\documentclass{PoS}

\usepackage{multirow}
\usepackage{txfonts}
\usepackage{xspace}
\usepackage{amsbsy}
\usepackage{amssymb}
\usepackage{IEEEtrantools}
\usepackage{setspace}

\title{Heavy Higgs boson production and decay into top quarks at the LHC}

\ShortTitle{Heavy Higgs boson production and decay into top quarks at the LHC}

\author{Werner Bernreuther$^a$, \speaker{Peter Galler}$^b$,
        Clemens Mellein$^a$, Zong-Guo Si$^c$ and Peter Uwer$^b$\\
        \llap{$^a$}Institut f\"ur Theoretische Teilchenphysik und Kosmologie,\\
        RWTH Aachen University, 52056 Aachen, Germany\\
        \llap{$^b$}Institut f\"ur Physik, Humboldt-Universit\"at zu Berlin,
        12489 Berlin, Germany\\
        \llap{$^c$}School of Physics, Shandong University, Jinan,
        Shandong 250100, China\\
        E-mail: \email{breuther@physik.rwth-aachen.de},
                \email{galler@physik.hu-berlin.de},
                \email{mellein@physik.rwth-aachen.de},
                \email{zgsi@sdu.edu.cn},
                \email{uwer@physik.hu-berlin.de}}

\abstract{
In this contribution we report on the calculation of the next-to-leading order
(NLO) QCD corrections to the hadro-production of heavy neutral Higgs bosons
and their decay into top-quark pairs within the type-II two-Higgs-doublet
extension of the standard model (SM). We take into account the contributions
from resonant Higgs boson production, the non-resonant SM \ttbar background
as well as the interference of these two contributions. The NLO corrections to
the signal and interference contributions are calculated by applying the heavy
top-quark mass ($m_t$) limit including an effective rescaling. In our NLO
calculation the QCD-Higgs interference is evaluated in the resonance region
that provides the dominant part of the heavy Higgs-boson contributions.
Evaluating representative $CP$-conserving and $CP$-violating parameter
scenarios within the two-Higgs-doublet model (2HDM) we present results for
different distributions and, in addition, for observables that depend on
the top-quark spin.
}

\FullConference{38th International Conference on High Energy Physics\\
		3-10 August 2016\\
		Chicago, USA}

\def\ttbar{\ensuremath{t\bar{t}}\xspace}
\def\mtt{\ensuremath{M_{t\bar{t}}}\xspace}

\def\fig#1{Fig.~\ref{#1}}

\def\tab#1{Tab.~\ref{#1}}
\def\vev{{\varv}}

\begin{document}
%
\section{Introduction\vspace{-0.4cm}}
\label{sec:intro}
Now that the standard-model like Higgs boson has been experimentally
\cite{Khachatryan:2014jba,Aad:2015gba} established at the LHC, the ongoing and
future investigation of the scalar sector---including the search for additional
spin-zero bosons---remains one of the central research goals at the LHC.
In this work, we investigate additional heavy Higgs bosons $\phi$ with masses
$m_{\phi}>2m_t$ and unsuppressed couplings to top quarks. Thus these heavy
Higgs bosons can appear as resonances in the \ttbar decay channel. So far
dedicated searches by ATLAS \cite{Aad:2015fna,ATLAS:2016pyq} and
CMS \cite{Chatrchyan:2013lca, Khachatryan:2015sma} do not show any excess above
the SM \ttbar continuum. However, until now the experimental analysis suffered from
the limited statistics of Run I and the fact that only leading-order (LO)
predictions for this process were available. Only very recently QCD NLO
corrections have been calculated \cite{Bernreuther:2015fts,Hespel:2016qaf}.\\
Here we report on the results of the computation of NLO QCD corrections to
heavy Higgs boson production and decay to top quarks presenting inclusive
as well as differential cross sections at a center of mass energy
$\sqrt{s}=13$TeV at the LHC. In addition, we study \ttbar spin correlations.
Our NLO calculation takes the following contributions into account: the
resonant production of heavy Higgs bosons and their decay into \ttbar,
the SM non-resonant contribution to \ttbar production and the interference
of these two processes. We apply the heavy top-quark mass limit to the
calculation of the NLO QCD corrections while at LO we keep the full $m_t$
dependence. An additional rescaling is used to improve the prediction
of this approximation. The calculation of the QCD-Higgs interference is
restricted to the resonant region which gives the dominant heavy
Higgs-boson contribution. Details of the calculation as well
as analytical results are presented in \cite{Bernreuther:2015fts}. 
Our approach can be applied to a broader spectrum of new physics models
featuring heavy spin-zero bosons that couple to \ttbar. However, we choose
a specific model to account for the decay width of the heavy Higgs bosons
in a consistent fashion because the effects of heavy Higgs bosons on
\ttbar production depend on the bosons' decay width. As a specific model,
we chose the UV-complete type-II 2HDM with no tree-level flavor-changing
neutral currents.
%
%
\section{Two-Higgs-doublet model scenarios\vspace{-0.4cm}}
\label{sec:scenarios}
We study the type-II 2HDM within the so-called alignment limit, i.e. the
lightest of the three neutral Higgs bosons is identified with the observed
125 GeV Higgs boson with SM-like couplings. Furthermore, we set the masses
$m_{2,3}$ of the other two heavy neutral Higgs bosons to $m_{2,3}>2m_t$.
In the next section, we present the results of four type-II 2HDM parameter
scenarios where the Higgs bosons have the same or slightly enhanced top-quark
Yukawa coupling strengths compared to the SM strength $m_t/\vev$. Scenarios 1, 2
and 4 are $CP$-conserving while scenario 3 is $CP$-violating. While in scenario 1
the two heavy Higgs bosons have similar masses, scenarios 2--4 feature
well-separated Higgs boson masses. A summary of all input parameters as well
as the Yukawa couplings and decay widths for the four scenarios is shown in
\tab{tab:scenarios}. For the definition of the parameters and further details
on scenarios 1--3 we refer to \cite{Bernreuther:2015fts}. Details of the 2HDM
in general and type-II in particular can be found in the
literature, e.g. \cite{Branco:2011iw}.
%
%
\begin{table}
\centering
{\scriptsize
\begin{tabular}{clcccc}
\hline
\hline
& & scenario 1 & scenario 2 & scenario 3 & scenario 4\\
\hline
\multicolumn{6}{l}{input parameters\vspace{0.1cm}}\\
&$\tan\beta$ & 0.7 & 0.7 & 0.7 & 1\\
&$\vev$ [GeV] & 246 & 246 & 246 & 246\\
&$m_+$ [GeV] & >720& >720 & >720 & >720\\
&$m_1$ [GeV] & 125 & 125 & 125 & 125\\
&$m_2$ [GeV] & 550 & 550 & 500 & 400\\
&$m_3$ [GeV] & 510 & 700 & 800 & 900\\
&$\alpha_1$ & $\beta$ & $\beta$ & $\beta$ & $\beta$\\
&$\alpha_2$ & 0 & 0 & $\pi/15$ & 0\\ 
&$\alpha_3$ & 0 & 0 & $\pi/4$ & 0\\ 
\hline
\multicolumn{6}{l}{calculated parameters\vspace{0.1cm}}\\
&$\Gamma_2$ [GeV] & 34.56 & 34.49 & 36.55 & 3.99\\
&$\Gamma_3$ [GeV] & 49.28 & 75.28 & 128.16 & 162.39\\
&$a_{t1}$ & 1 & 1 & 0.978 & 1\\
&$b_{t1}$ & 0 & 0 & 0.297 & 0\\
&$a_{t2}$ & 1.429 & 1.429 & 0.863 & 1\\
&$b_{t2}$ & 0 & 0 & 0.988 & 0\\
&$a_{t3}$ & 0 & 0 & -1.157 & 0\\
&$b_{t3}$ & 1.429 & 1.429 & 0.988 & 1\\
\hline
\hline
\end{tabular}}
\caption{2HDM parameter settings for scenarios 1--4. The SM-like Higgs boson
$\phi_1$ has a decay width $\Gamma_1\approx 4$~MeV. It plays no role in our
analysis.}
\label{tab:scenarios}
\end{table}
%
%
%
\section{Results\vspace{-0.4cm}}
\label{sec:results}
\begin{figure}
\includegraphics[scale=0.6]{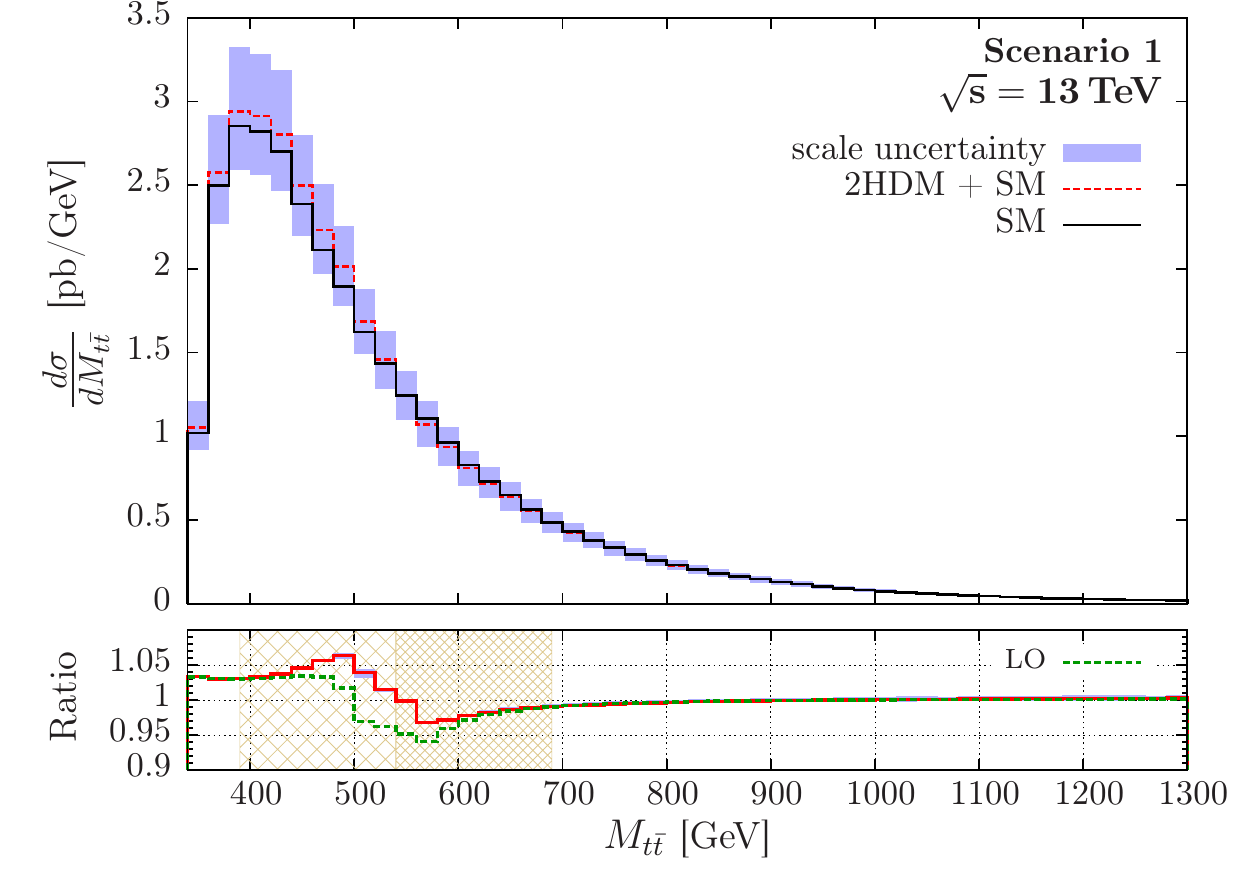}
\includegraphics[scale=0.6]{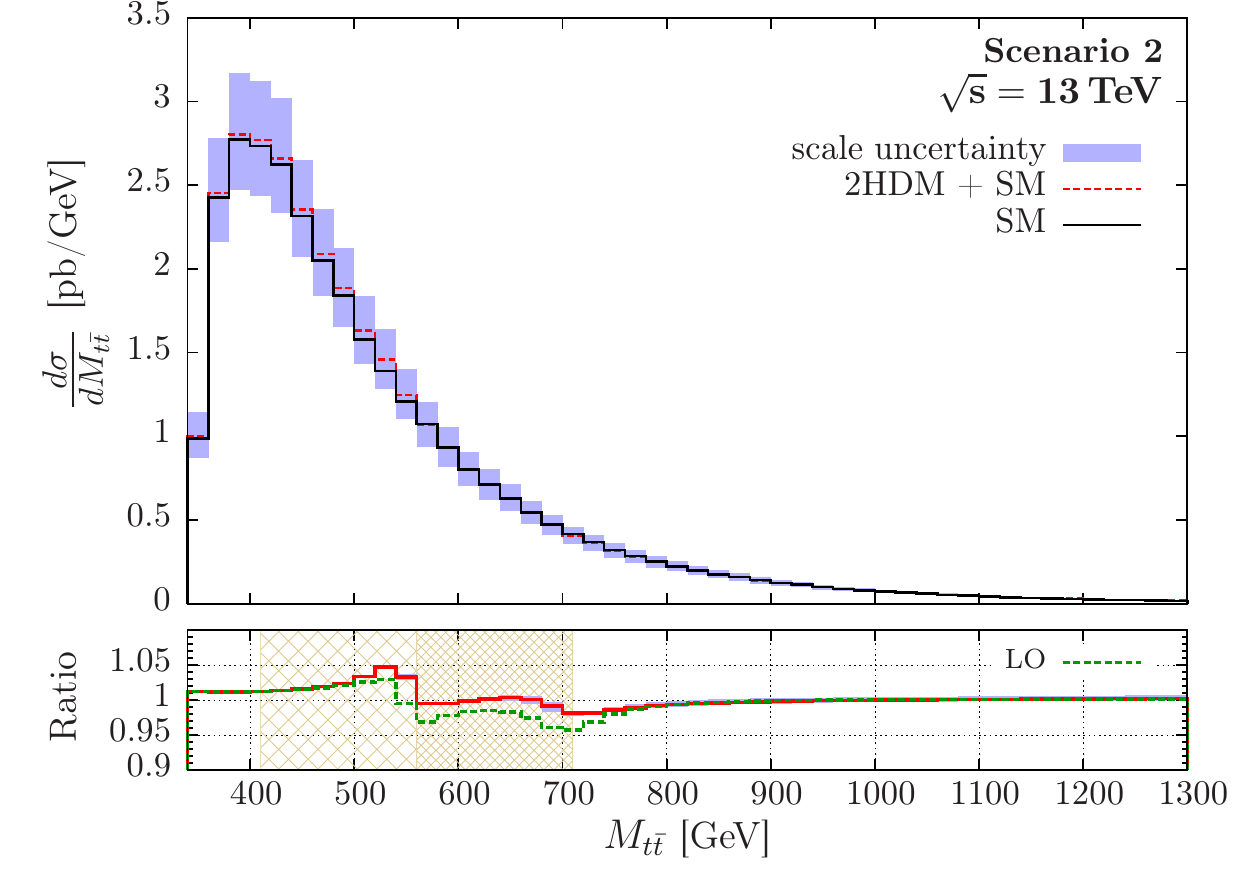}
\includegraphics[scale=0.6]{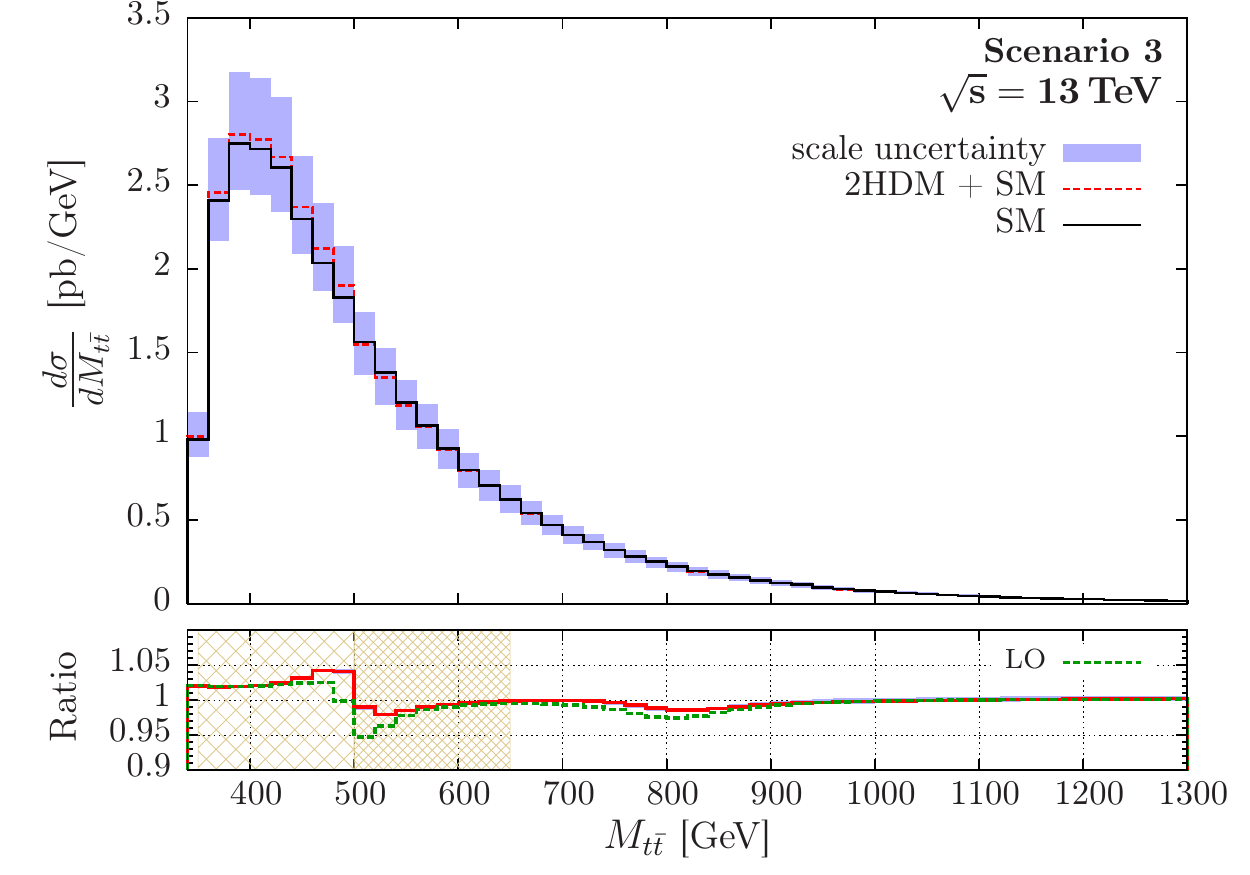}
\includegraphics[scale=0.6]{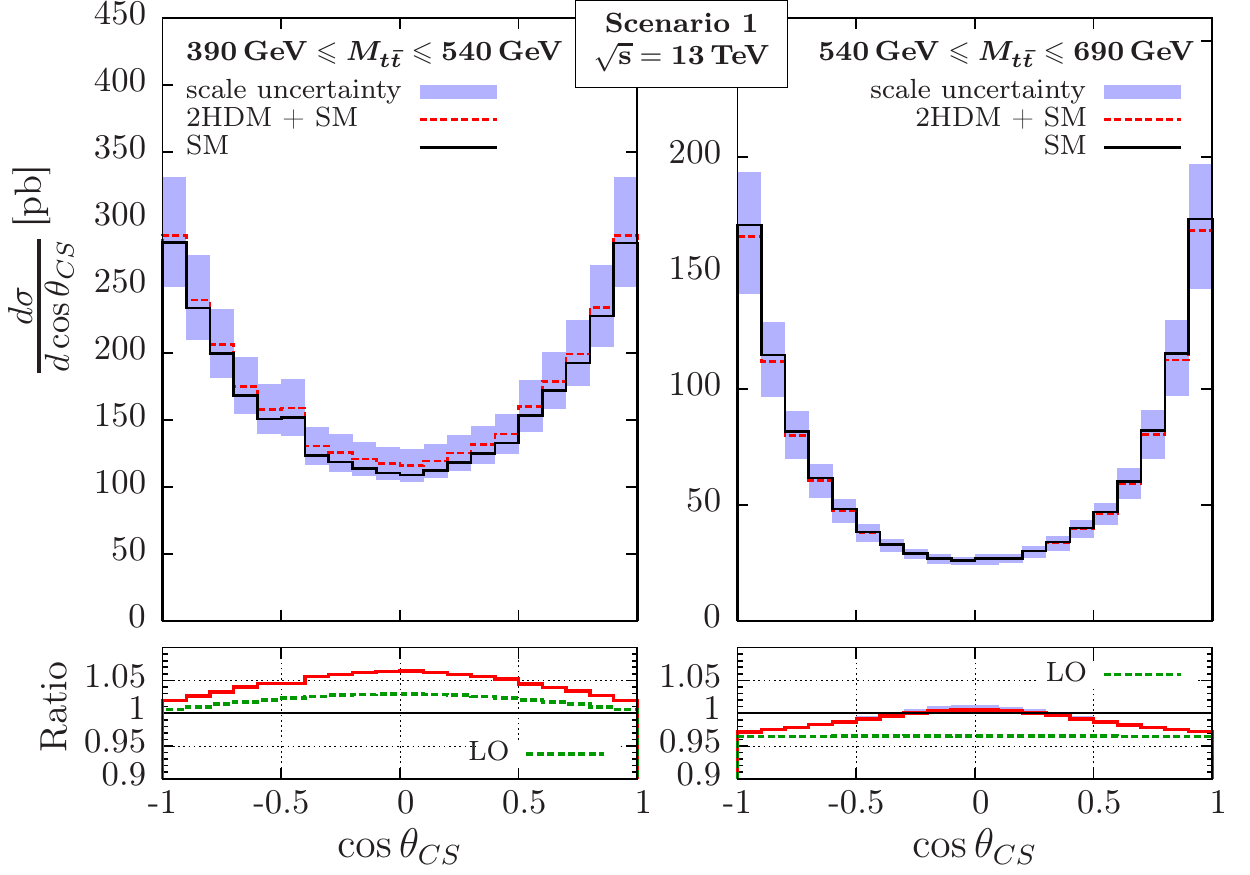}
\caption{
{\bf Upper left:} \mtt distribution for scenario 1,
{\bf upper right:} \mtt distribution for scenario 2,
{\bf lower left:} \mtt distribution for scenario 3,
{\bf lower right:} Collins-Soper angle distribution for scenario 1
in two different \mtt bins.
}
\label{fig:Mtt}
\end{figure}

\begin{figure}
\begin{center}
\includegraphics[scale=0.9]{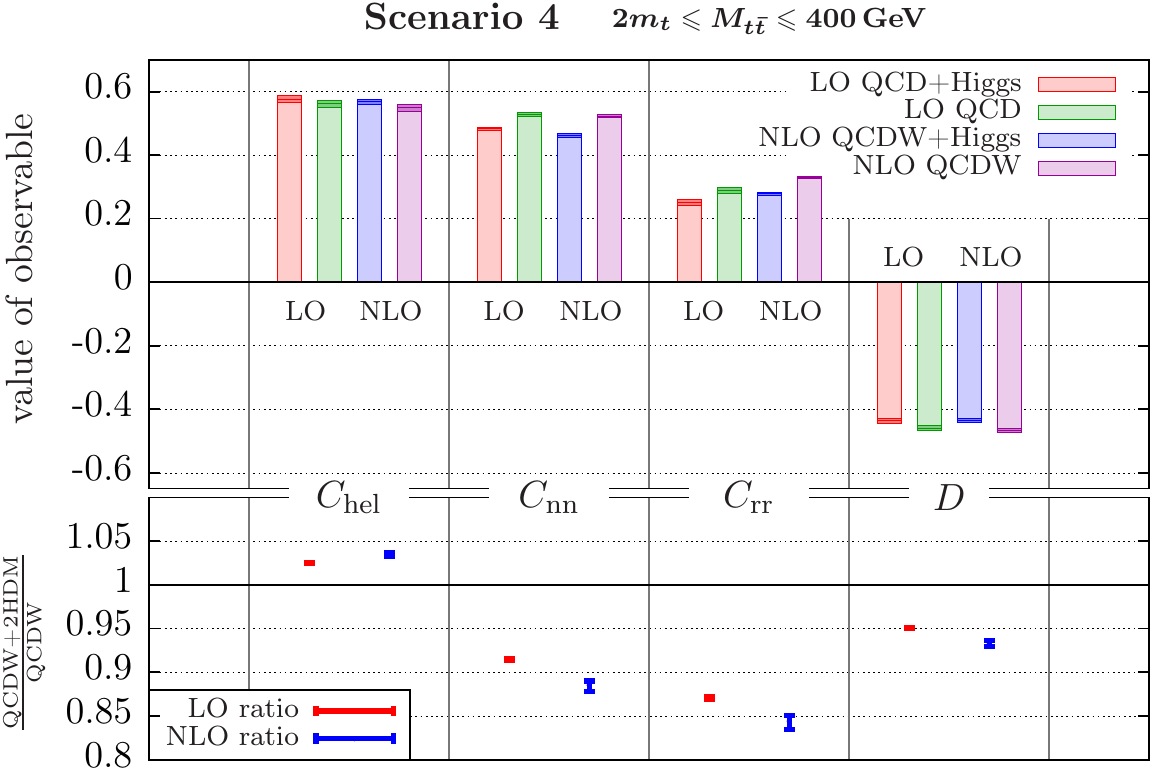}
\end{center}
\vspace{-0.5cm}
\caption{Spin correlations for scenario 4 at LO and NLO. Upper plot: values
of the observables, lower plot: signal-to-background ratios.\vspace{-0.5cm}}
\label{fig:spin}
\end{figure}

In this section we present results for the NLO QCD corrections to heavy Higgs
production and decay to \ttbar within the type-II 2HDM using the two
approximations (heavy-top limit and restriction to resonance region) mentioned
in the introduction.
%
%
\paragraph{Spin independent observables:}
We calculated, for scenarios 1--3, a number of observables that do not depend
on the top-quark spin. The inclusive \ttbar cross section is shown in \tab{tab:inclxs}.
The central value of the renormalization and factorization scales
is denoted by $\mu_0=(m_2+m_3)/4$. The scales are changed simultaneously
by a factor of 2 and $1/2$, respectively. The resulting change in the total
cross section is indicated in this table by the superscripts and subscripts.
$\sigma_{\rm{QCDW}}$ is the \ttbar cross section due to QCD and weak
interaction at NLO and $\sigma_{\rm{2HDM}}$ denotes the contributions from
resonant heavy Higgs production and decay to \ttbar and the interference with
the non-resonant SM \ttbar production at NLO QCD.
Table \ref{tab:inclxs} shows that the ratio of the 2HDM and the SM
contribution is only about 1--2\%. A higher sensitivity can be achieved by
studying differential cross sections. In \fig{fig:Mtt} we present results for
the top quark pair invariant mass \mtt distribution for scenarios 1--3 at
NLO QCD. In the upper plots, the SM contribution and the SM+2HDM contribution
including the interference is shown in black and red, respectively. The blue
shaded area indicates the uncertainty due to renormalization and factorization
scale variations. In the lower plots, the signal-to-background ratio at LO
(NLO) is displayed in green (red). The lower-lying Higgs resonance leads to
a peak-dip structure in the \mtt distribution that is caused by the
interference of signal and background. In the inclusive \ttbar cross section
the peak and dip partly cancel. As a result the effects of the additional
Higgs bosons are significantly reduced as the numbers of \tab{tab:inclxs} show.
For avoiding this cancellation we applied cuts on \mtt to restrict the
observables to a regime below and above the resonance. This is indicated by
the hatched region in the ratio plots in \fig{fig:Mtt}.
We investigated, with these \mtt bins, the differential cross section with 
respect to the top-quark rapidity, the top-quark transverse momentum and the
cosine of the Collins-Soper angle $\cos\theta_{\rm{CS}}$. Among these
observables the Collins-Soper angle is the most sensitive one. We show the
result in \fig{fig:Mtt} only for scenario 1 where the effects are largest due
to the overlapping resonances. The maximum signal-to-background (S/B) ratios in
the \mtt distribution and Collins-Soper angle distribution in the lower \mtt
bin are of similar size ($\gtrsim 6\%$). But in the case of the \mtt
distribution one needs a small bin size to reach this sensitivity. Considering
the inclusive cross section in the range $390\rm{GeV}\le\mtt\le 540\rm{GeV}$
one obtains a reduced value for S/B of $\sim 4\%$. This also shows that
measurements of the \mtt distribution with a large bin size tend to wash out
the effects of additional heavy Higgs bosons in \ttbar production.
%
%
\begin{table}[h!]
\centering
{\scriptsize
{\renewcommand{\arraystretch}{1.2}
\renewcommand{\tabcolsep}{0.2cm}
\begin{tabular}{llll}
\hline
\hline
& Scenario 1 & Scenario 2 & Scenario 3\\
\hline
$\mu_0$ [GeV]& 265 & 312.5 & 325\\
$\sigma_{\rm{QCDW}}$ [pb] & $643.22^{+81.23}_{-77.71}$&
$624.25^{+80.98}_{-76.19}$& $619.56^{+81.05}_{-75.72}$\\
$\sigma_{\rm{2HDM}}$ [pb] & $13.59^{+1.85}_{-1.64}$&
$7.4^{+0.77}_{-0.78}$& $7.21^{+0.81}_{-0.77}$\\
$\sigma_{\rm{2HDM}}/\sigma_{\rm{QCDW}}$ [\%]& 2.1& 1.2& 1.2\\
\hline
\hline
\end{tabular}}}
\caption{Inclusive \ttbar cross sections at NLO QCD with and without the heavy
Higgs resonances.}
\label{tab:inclxs}
\end{table}
%
%
%
%
%
%
%
\paragraph{Spin correlations:}
We investigated also the polarization of the top quark and \ttbar spin
correlations for several type-II 2HDM scenarios. Here, we present only the
results for \ttbar spin correlations in scenario 4 and refer for a detailed
account of spin dependent observables to a future publication
\cite{Bernreuther:2016}. We calculated four spin correlation observables
$C_{\rm{kk}}\equiv C_{\rm{hel}}$, $C_{\rm{nn}}$, $C_{\rm{rr}}$ and
\mbox{$D=-(C_{\rm{hel}}+C_{\rm{nn}}+C_{\rm{rr}})/3$} within appropriate \mtt bins
in order to enhance the signal. These spin correlations are equivalent to
angular correlations between the leptonic decay products of the top quark:
\begin{equation}
C_{\rm{xx}}=-9\langle\cos\theta_{x,+}\cos\theta_{x,-}\rangle,
\quad D=-3\langle\cos\theta_{+,-}\rangle,
\quad\cos\theta_{x,\pm}=\hat{\mathbf{x}}\cdot\hat{\boldsymbol{\ell}}_{\pm},
\quad\cos\theta_{+,-}=\hat{\boldsymbol{\ell}}_+\cdot\hat{\boldsymbol{\ell}}_-,
\end{equation}
where $\langle\mathcal{O}\rangle=\sigma^{-1}\int d\sigma\mathcal{O}$
denotes the expectation value of the observable $\mathcal{O}$ and
$\hat{\mathbf{x}}=\{\hat{\mathbf{k}},\hat{\mathbf{n}},\hat{\mathbf{r}}\}$.
Here $\hat{\mathbf{k}}$ is the direction of flight of the top quark in the
\ttbar zero-momentum frame, $\hat{\boldsymbol{\ell}}_{\mp}$ is the direction of
flight of the lepton $\ell_-$ (anti-lepton $\ell_+$) in the anti-top (top)
rest frame. The vectors $\hat{\mathbf{n}}$ and $\hat{\mathbf{r}}$ are defined as
\begin{equation}
\hat{\mathbf{n}}=\frac{\hat{\mathbf{p}}\times\hat{\mathbf{k}}}
{|\hat{\mathbf{p}}\times\hat{\mathbf{k}}|},\quad
\hat{\mathbf{r}}=\frac{\hat{\mathbf{p}}-y\hat{\mathbf{k}}}
{|\hat{\mathbf{p}}-y\hat{\mathbf{k}}|},\quad
y=\hat{\mathbf{p}}\cdot\hat{\mathbf{k}}
\end{equation}
and $\hat{\mathbf{p}}$ is the direction of flight of the incoming proton in the
laboratory frame.\\
In \fig{fig:spin} we display the LO and NLO QCD results for the four different
spin correlations in scenario~4 for $2m_t\le\mtt\le 400\rm{GeV}$. The upper
plot shows the values of the spin correlations at LO and NLO for the SM
contribution and the combined SM and 2HDM contributions. The darker parts of the
bars in the upper plot indicate the uncertainties due to renormalization and
factorization scale variations. In the lower plot the corresponding ratios
between the 2HDM contributions and the SM contributions are shown. The
error bars indicate the change due to scale variations. It seems that the NLO
results show a stronger dependence on the renormalization and factorization
scales than the LO even though one would expect the opposite behavior. At
LO the renormalization scale dependence cancels in the ratio. As a
consequence the variation of $\mu_0$ does not give a reliable estimate of
the uncertainty of the LO result. However, the NLO corrections affect the
ratios only slightly and increase their absolute values by a few percent.\\
The largest effect of heavy Higgs bosons in \ttbar production can be
seen in $C_{\rm{rr}}$, which shows a deviation of $\sim 15\%$ from its SM
value. This is to be compared with the signal-to-background ratio of
$\sim 4\%$ for the cross section in the same \mtt bin. It clearly shows that spin
dependent observables such as spin correlations significantly increase the
sensitivity to heavy Higgs bosons.
%
%
%
%
\newpage
%
\section{Conclusion\vspace{-0.4cm}}
\label{sec:concl}
We calculated, within the type-II 2HDM extension of the SM, the NLO QCD
corrections to the resonant production of heavy Higgs bosons and their decay
into top-quark pairs at the LHC, taking into account the non-resonant SM \ttbar
continuum and its interference with the signal. We studied $CP$-conserving and
$CP$-violating scenarios and investigated several observables. The interference
contribution generates a peak-dip structure that leads to cancellations in
inclusive observables. As a consequence the total cross sections for our 2HDM
scenarios show only small signal-to-background ratios of about 1--2\%.
Differential distributions in combination with cuts on \mtt can give larger
effects. In scenario 1 where the two resonances overlap we calculated a
signal-to-background ratio of up to $6\%$ for the distributions of \mtt and
the Collins-Soper angle. While this is already an increase of S/B by a factor
of three, less optimistic parameter scenarios call for a higher sensitivity.
We showed that this can be achieved by studying \ttbar spin correlations.
As an example we presented the results for a type-II 2HDM scenario that contains
two heavy Higgs bosons with masses of 400 GeV and 900 GeV and Yukawa couplings
to top quarks of SM strength. Within this scenario the spin correlation
$C_{\rm{rr}}$ shows the largest signal-to-background ratio of about 15\%. In
comparison with the signal-to-background ratio of $\sim 4\%$ of the total
cross section evaluated in the same \mtt bin as $C_{\rm{rr}}$ this amounts
to an increase of about a factor of four in sensitivity. In this analysis we
have shown that the challenges of searching for heavy Higgs resonances in the
\ttbar decay channel can partly be overcome by a combination of top-spin
independent and spin-dependent observables evaluated in suitable \mtt windows.
%
%
\begin{spacing}{0}
\bibliographystyle{JHEP}
\bibliography{references}
\end{spacing}
\end{document}